\documentclass[aps,prl,twocolumn,superscriptaddress
]{revtex4-2}
\usepackage{chngcntr}
\usepackage{graphicx}
\usepackage{color}
\usepackage{amsmath}
\usepackage{enumitem}
\usepackage{amssymb}
\usepackage{hyperref}
\usepackage{cancel}
\usepackage{ulem}
\usepackage{multirow}
\usepackage{CJK}
\usepackage{verbatim}
\usepackage{dcolumn}
\usepackage{bm}
\usepackage{subfigure}
\usepackage{xcolor,float}

\newcommand{\be}{\begin{equation}}
	\newcommand{\ee}{\end{equation}}

\newcommand{\bea}{\begin{eqnarray}}
	\newcommand{\eea}{\end{eqnarray}}

\newcommand{\lp}{\left(}
\newcommand{\rp}{\right)}

\renewcommand{\vec}[1]{{\boldsymbol #1}}

\newcommand\startsupplement{%
       \newpage\clearpage
       \setcounter{secnumdepth}{2}
       \setcounter{table}{0}
       \renewcommand{\thetable}{S\arabic{table}}
       \setcounter{figure}{0}
       \renewcommand{\thefigure}{S\arabic{figure}}
       \setcounter{equation}{0}
       \renewcommand{\theequation}{S\arabic{equation}}
       \setcounter{section}{0}
       \renewcommand{\thesection}{Section \Roman{section}}
       \renewcommand{\thesubsection}{\Roman{section}. \Alph{subsection}}
    }
\begin{document}
		
\title{Numerical evidence for spin chirality emerging from itinerant ferromagnets in bands with Berry curvature}

\author{Shuai Yang}	
\affiliation{Department of Physics and State Key Laboratory of Surface Physics, Fudan University, Shanghai 200433, P.R. China}
\author{Zhiyu Dong}
\affiliation{Department of Physics and Institute for Quantum Information and Matter, California Institute of Technology, Pasadena, California 91125, USA}
\affiliation{Department of Physics, Massachusetts Institute of Technology, Cambridge, MA 02139, USA}
\author{Yan Chen}
\email{yanchen99@fudan.edu.cn}
\affiliation{Department of Physics and State Key Laboratory of Surface Physics, Fudan University, Shanghai 200433, P.R. China}
\affiliation{Shanghai Branch, Hefei National Laboratory, Shanghai 201315, P.R. China}

\date{\today}

\begin{abstract}
The behavior of strongly interacting electrons in bands with Berry curvature is a problem of wide interest. In this paper, we study this problem by numerically studying a fluxed Hubbard-type model on square lattice. Using this model, we demonstrate a metallic ferromagnet in electron bands equipped with Berry curvature can develop non-coplanar spin order in which spin polarization axes at different position span finite solid angles. We find spin chirality can emerge in this setting by doping or adding gauge flux on top of a collinear ferromagnet. This result supports the prediction of spin chirality occurring through an emergent spin orbital interaction.
Meanwhile, our result shows that, on top a ferromagnetic background, the spin chirality emerges at a finite threshold value of orbital magnetization, resembling the predicted behavior in theory. 
\end{abstract}
	
\maketitle
	
The Berry phase\cite{Berry1984QuantalPF}, a geometric phase accumulated by a quantum state during adiabatic evolution along a closed loop in parameter space, profoundly affects the behavior of electrons in quantum materials. It can originate in either
momentum $\boldsymbol{k}$-space or real $\boldsymbol{r}$-space
. In $\boldsymbol{k}$-space, the Berry phase can induce chiral motion of electrons, leading to various phenomena such as anomalous Hall conductivity \cite{RevModPhys.82.1539}, orbital magnetizations (OM)\cite{RevModPhys.82.1959, PhysRevLett.95.137205}, and topological insulators\cite{RevModPhys.83.1057, RevModPhys.82.3045}. In $\boldsymbol{r}$-space, the Berry phase can be detected by carriers moving through non-coplanar magnetic textures, resulting in observable signatures like Hall conductivity\cite{PhysRevB.62.R6065, PhysRevLett.87.116801, Nagaosa_2012,10.1063/1.3665219}.

An intriguing scenario arises when both types of Berry phase coexist within a system, enabling carriers to simultaneously perceive $\boldsymbol{r}$-space and $\boldsymbol{k}$-space Berry phases. This situation is potentially realized in recently discovered ferromagnetism (FM) in graphene systems\cite{cao2018correlated,cao2018unconventional,zhou2021half,zondiner2020cascade,andrei2020graphene,seiler2022quantum}. These systems serve as a realistic platform that encompasses both components: they host Dirac bands, featuring a significant $\boldsymbol{k}$-space Berry curvature\cite{andrei2020graphene}. Concurrently, FM offers the potential to develop a non-coplanar spin texture, giving rise to the $\boldsymbol{r}$-space Berry phase.

\begin{figure}
	\includegraphics[scale=0.42]{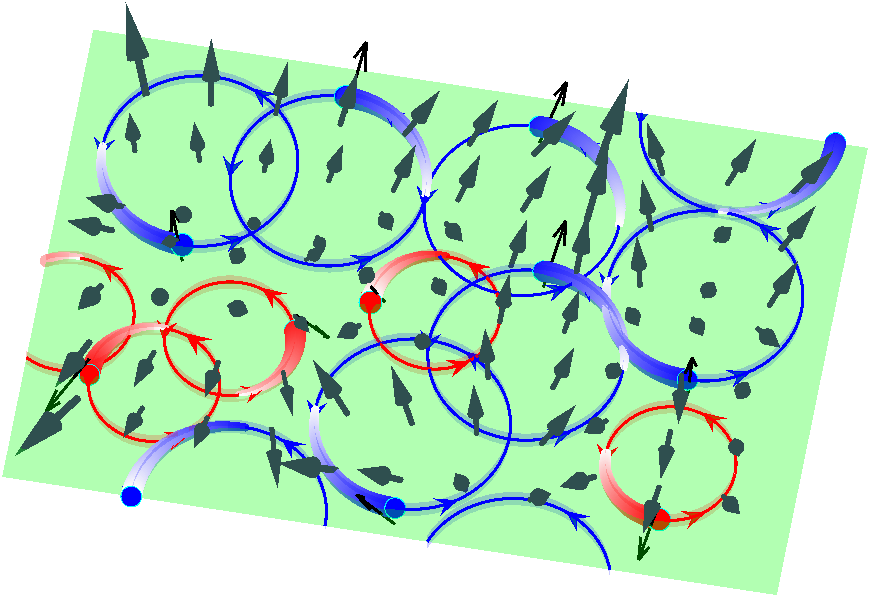}
	\caption{
 A cartoon illustrating the spin chirality driven by the emergent SOI. Blue and red balls represent the majority-spin and minority-spin carriers, respectively. The circular trajectories depict their chiral motion due to the band's Berry curvature. The size of the trajectories represents the orbital magnetization of two types of carriers $\mathcal{M}_+$ and $\mathcal{M}_-$, which are unequal. It gives rise to the emergent SOI (see Eq.\ref{eq:H_SOI}), leading to a textured mean-field spin polarization, indicated by the black arrows. }
	\label{schematic}
\end{figure}

Recent theoretical studies shed light on this issue, particularly those referenced in\cite{dong2022chiral, PhysRevLett.130.206701}, which examined the behavior of electrons in itinerant ferromagnets with band Berry curvature. These works reveal that the orbital motion of carriers gives rise to an emergent spin-orbit interaction (SOI), formulated as follows:
\begin{equation}
    \begin{aligned}
        &\mathcal{H}_{SOI} = -\lp \mathcal{M}_+ - \mathcal{M}_-\rp B,\\
        &B = \frac{\phi_0}{4\pi }\boldsymbol{S}\cdot(\partial_x \boldsymbol{S}\times\partial_y\boldsymbol{S}),\quad \phi_0=\frac{hc}{e}
    \end{aligned}
    \label{eq:H_SOI}
\end{equation}
where $\vec S$ is a unit vector field representing the direction of the spatial-dependent spin polarization in ferromagnets, $ \mathcal{M}_+$ and $ \mathcal{M}_-$ denote the OM of the majority-spin and minority-spin carriers. Distinct from the conventional microscopic SOI, this emergent SOI respects the $SU(2)$ spin rotation symmetry. Consequently, it modifies the typical collinear spin order observed in itinerant ferromagnets, transforming it into a noncoplanar spin texture with a nonvanishing chirality, as depicted in Fig. \ref{schematic}.
This way of generating chiral spin texture is distinct from those previously studied in Refs.\cite{PhysRevLett.101.156402, PhysRevLett.108.096403, PhysRevLett.118.147205, PhysRevLett.109.166405}. These previous studies rely on the nesting of Fermi surfaces to form triple-$Q$ magnetic orders. In contrast, this mechanism requires neither specific nested wave vectors nor a magnetic coupling between itinerant electrons and local magnetic moments. The resulting spin chirality arises from the FM instability of the itinerant electrons and is driven by the OM. 

In this paper, we provide numerical evidence supporting the existence of this interaction as well as its effects. Namely, we employ a minimal microscopic lattice model that hosts itinerant FM and band with Berry curvature. Using the density-matrix-renormalization-group (DMRG)\cite{itensor} approach, we investigate the ground state of this model and observe that the itinerant electrons exhibit chiral magnetic order.
Moreover, we show that the spin chirality can be switched on and off upon switching the orbital magnetization. This further verifies that the observed spin chirality originates from the emergent SOI proposed in Ref.\cite{dong2022chiral}.
	
\textit{Model Hamiltonian and Methodology.---}
To start, we look for a minimal 2D model that can capture the physics of emergent SOI. This model needs a magnetic flux to achieve time-reversal symmetry breaking, which will give rise to non-zero OM which is the predicted strength of emergent SOI. Meanwhile, this model has to break all reflection symmetry along any mirrors perpendicular to the 2D plane. For that, we need a two-sublattice system with inequivalence between them. Furthermore, as the emergent SOI we aim to test is predicted for spin-polarized state\cite{dong2022chiral}, we need to generate FM in this model. This can be achieved by introducing a strong Coulomb repulsion. 

\begin{figure}
	\includegraphics[scale=0.35]{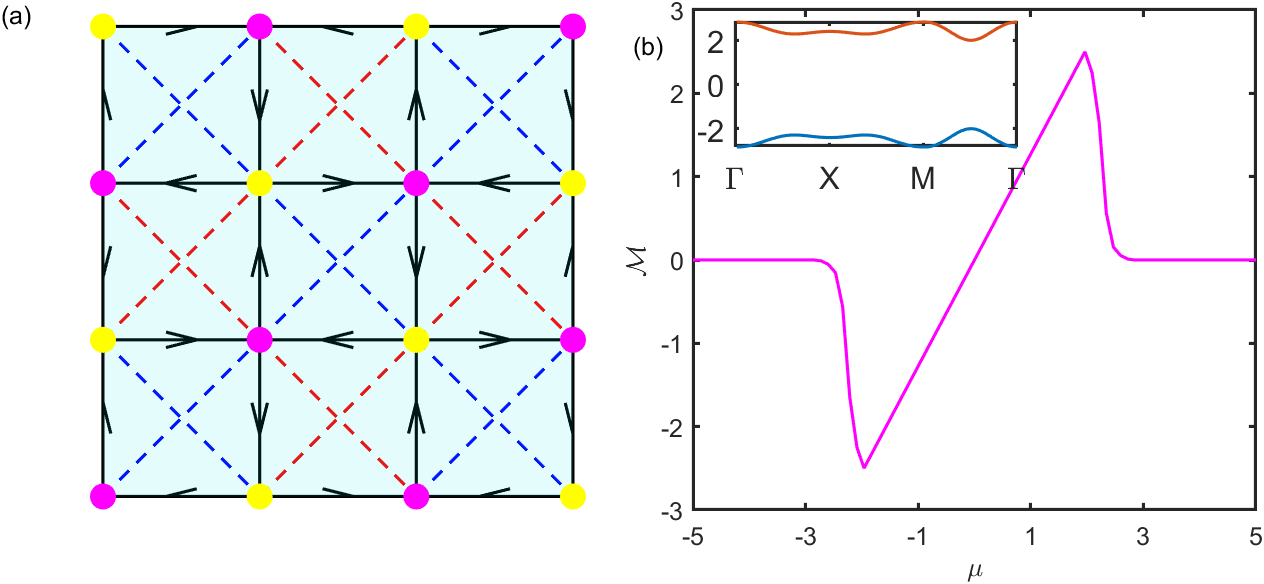}
	\caption{
 (a) Schematic of staggered flux square lattice Hubbard model, arrows on NN bond mark the direction of positive phase hopping, blue(red) NNN bond correspond $+(-)t_2$ hopping amplitude.
 (b)Total OM of one certain spin electron as a function of the chemical potential $\mu$ at $t_2/t_1=0.6,\phi=\pi/4$. Along the high symmetry points $\Gamma(0,0)-X(\pi,0)-M(\pi,\pi)-\Gamma(0,0)$ in the first Brillouin Zone, the corresponding band structure of the single particle Hamiltonian is presented with the inset. Each band is spin degenerate.
 }
	\label{Rs_and_OM}
\end{figure}

One model that hosts all these features is a square-lattice extended Hubbard model with gauge flux, which has been introduced and investigated in Ref.\cite{PhysRevB.99.014407,PhysRevB.92.245124}. In this paper, we show that this model indeed hosts an emergent SOI like the one proposed in Ref.\cite{dong2022chiral}. For convenience, we write down the model as follows: 
	\begin{equation}
		\begin{aligned}
			\mathcal{H}=&-t_1\sum_{\left<jm\right>,\sigma}\left(e^{i\phi\delta_1^{jm}}c_{j\sigma}^{\dagger}c_{m\sigma}+H.c.\right)\\
   &-t_2\sum_{\left<\left<jm\right>\right>,\sigma}\left({\delta_2^{jm}}c_{j\sigma}^{\dagger}c_{m\sigma}+H.c.\right)\\
			&+U\sum_{j}n_{j\uparrow}n_{j\downarrow}
		\end{aligned}
	\end{equation}
where $c_{j\sigma}^\dagger$ and $c_{j\sigma}$ are
the creation and annihilation operators of an electron with spin $\sigma={\uparrow,\downarrow}$ at site $j$, respectively. The operators $n_{j\uparrow}(n_{j\downarrow})$ count the number of particles with spin $\uparrow(\downarrow)$ at site $j$. The parameter $U$ denotes the on-site Coulomb interaction strength. Here, $\left<\cdots\right>,\left<\left<\cdots\right>\right>$ represent the nearest-neighbor(NN) and next-nearest-neighbor(NNN) pairs of sites. And the quantities $t_1$ and $t_2$ are the magnitudes of NN and NNN hoppings, respectively.
For the NN hopping amplitude between site $j$ and $m$, we assign $\delta_1^{jm}=+(-)$ depending on whether the electron hops in the direction (reverse direction) of solid black arrow depicted in Fig. \ref{Rs_and_OM}(a). As for the NNN hopping, $\delta_2^{jm}$ takes a value of $+1$ when NNN bond $jm$ aligns with the blue lines in Fig.\ref{Rs_and_OM}(a), and $-1$ when it aligns with the red lines.
The staggered NNN hopping is necessary to break the mirror symmetries along the $x$ and $y$ direction, otherwise, the OM would be suppressed due to symmetry constraint.

Given the pivotal role of OM in governing electron behavior in our system, we calculate the value of $\mathcal{M}$ for the single-particle Hamiltonian using the standard results in Ref.\cite{PhysRevLett.97.026603}\footnote{calculation and results on orbital magnetization are detailed in the supplementary materials.}. 
The value of OM as a function of the chemical potential $\mu$ is illustrated in
Fig. \ref{Rs_and_OM}(b). Here the calculation is done at the parameter of 
 $t_2/t_1=0.6,\phi=\pi/4$.
As the carrier density increases, the absolute value of $\mathcal{M}$ increases until 
the Fermi level intersects the valence band. When $\mu$ lies inside the band gap, the OM exhibits a linear dependence on $\mu$. The slope is proportional to the Chern number of the occupied bands, as expected in Ref.\cite{xiao2010berry}.

This model features a nonvanishing OM at a generic value of $\phi$. 
However, the OM is expected to vanish at $\phi=0,\pi/2,\pi$. This can be understood through the following symmetry analysis: At $\phi=0$, the Hamiltonian preserves time-reversal symmetry, thus dictating a vanishing OM. At $\phi=\pi$, the total flux threading through each triangular plaquette is $2\pi$, which restores the time-reversal symmetry. At $\phi=\pi/2$, the model breaks mirror symmetry along the direction of the NN-bond but still preserves mirror symmetry along NNN direction and time-reversal symmetry. This occurs because electrons perceive a flux of $\pi$ in each triangular plaquette. A flux of $\pi$ per plaquette respects both time-reversal symmetry and mirror reflection along the NNN-bond direction, as both operators transform the flux to $-\pi$-flux, which is equivalent to $\pi$-flux since a gauge transformation can absorb a flux difference of $2\pi$ in each plaquette. 

Below we test the theory of emergent SOI in several ways.
The first way is to show the presence of spin chirality, which is the main manifestation of the emergent SOI predicted in Ref.\cite{dong2022chiral}.
Another way to probe the presence of emergent SOI is by studying how spin chirality evolve upon tuning OM. We will do this ``experiment" by tuning the flux $\phi$. From theory we expect the spin chirality to be switched off as OM becomes smaller than a finite threshold value so that the chiral spin texture's energy gain from emergent SOI can no longer overcome the energy cost from spin stiffness\cite{dong2022chiral}. We will use numerics to show this is indeed the case.


In this model, it is known that one regime where a FM can be realized is when the band becomes flat. This is achieved at a fine-tuned parameter with $t_2/t_1=0.6,\phi=\pi/4$ when interaction strength $U$ is above a threshold\cite{PhysRevB.99.014407}.
Here, we take $U=4.0$ as an example. We are interested in the impact of orbital magnetization, therefore, we scan doping $\nu$, which effectively tunes the orbital magnetization of the carriers. We expect, upon scanning doping, system could develop chiral magnetism. To test this idea, we need to identify a) whether it is a ferromagnet, b) whether it has finite spin chirality.

We identify ferromagnetism through measuring the total spin of the system, which is defined as
\begin{equation}\label{eq:S^2}
    \frac{S^2}{S_{\rm{max}}^2}=\frac{1}{N_sS_{\rm_{max}}^2}\sum_{ij}\left<\boldsymbol{S_i}\cdot\boldsymbol{S_j}\right>,
\end{equation}
Here, $N_s$ is the system size, and $S_{\text{max}}^2 = \frac{N}{2}(\frac{N}{2}+1)$ represents the square of total spin in a fully polarized ferromagnet. In the summation, site indices $ij$ are summed over all sites.
When $S^2=0$, the system is a paramagnet without long-range FM order.  When $S^2$ is finite, the system is an FM. In numerics, we find for $\nu<0.42$, $S^2/S_{\text{max}}^2$ is nonzero, therefore, the system has FM order in this regime. In comparison, for $0.42<\nu < 0.5$, $S^2/S_{\text{max}}^2$ is on the order of less than \(10^{-7}\), which we identify as 0 since it is below the error generated by truncation $10^{-5}$ . In conclusion, ferromagnetism is suppressed in the regime of $0.42<\nu < 0.5$.

\begin{figure}
		\includegraphics[scale=0.5]{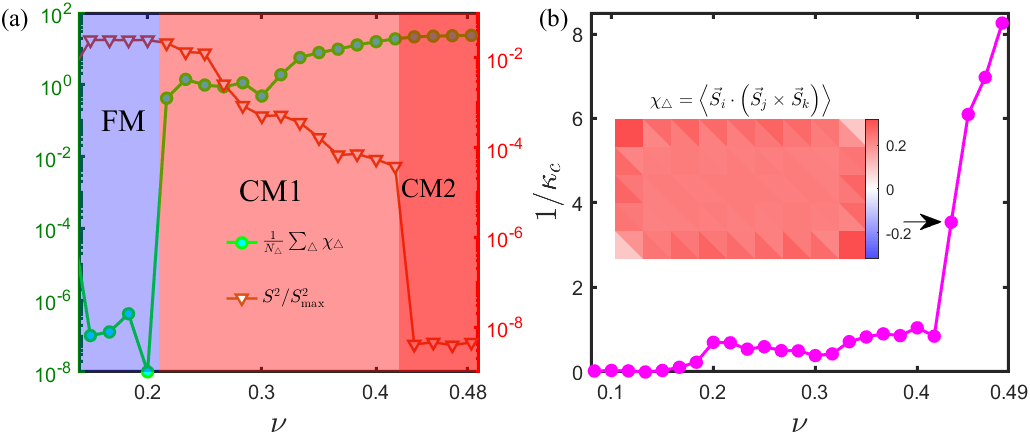}
		\caption{
        (a)Mean spin chirality $\frac{1}{N_\triangle}\sum_\triangle \chi_\triangle$and static spin structure factor $S_{\boldsymbol{k}}(k_x,k_y)$ on high-symmetric points of multi-leg ladders(both open boundary condition(OBC) in $x,y$ direction) with $N_x=10,N_y=6$  as a function of filling factor $\nu$, (b)Inverse charge compressibility $\kappa_c^{-1}\approx \frac{N_e^2}{N_s}[E(N_e+\delta N_e)+E(N_e-\delta N_e)-2E(N_e)]/\delta N_e^2$(where $E(N_e)$ is the ground-state energy with $N_e$ electrons on $N_s=N_x\times N_y$ sites and $\delta N_e^2=2^2$) with different filling factor $\nu$, there is a discontinuity when approaching the half-filling case.
The inset shows the pattern of discrete spin chirality $\left<\boldsymbol{S}_i\cdot(\boldsymbol{S}_j\times\boldsymbol{S}_k)\right>$  in the ground state. This result was measured under the parameters indicated by the black arrow in (b). Here, $ijk$ are chosen such that the loop $i\rightarrow j\rightarrow k\rightarrow i$ winds in the clockwise direction.
  }
		\label{kappac_and_mkHS} 
	\end{figure}

Additionally, upon increasing carrier density, 
we found the value of spin chirality $\chi_\Delta$ (shown by green solid circles in Fig.\ref{kappac_and_mkHS}(a)) differs by several orders of magnitude. Namely, the chirality is above $\mathcal{O}(10^{-2})$ in $\nu>0.2$, whereas at $\nu<0.2$, the chirality is below $10^{-5}$. 
We identify chirality on small-$\nu$ regime as 0 because it is below the truncation error $\epsilon\sim 10^{-7}$. 

In summary, the two observables, spin chirality and total spin $S^2$, enable us to identify a cascade of phases in this system: 
\begin{enumerate}
    \item[(i)] In the low-doping regime \(0.1 < \nu < 0.2\) is a collinear ferromagnetic phase (FM) 
which features finite $S^2$ and a vanishing spin chirality.
\item[(ii)] In intermediate-doping regime \(0.2 < \nu < 0.42\), which exhibits finite spin chirality along with magnetic long-range order (finite $S^2$). We call this phase type-I chiral metal (CM1).
\item[(iii)] In the high-doping regime \(\nu > 0.42\) spin chirality becomes so strong that the ferromagnetic order is completely suppressed, as signified by $S^2=0$. We label it as type-II chiral metal (CM2).
\end{enumerate}

All these three phases are metallic. We verify this by measuring the inverse compressibility $\kappa_c^{-1}=\frac{\partial \mu}{\partial n}$ (see Fig.\ref{kappac_and_mkHS}(b)). 
The inverse compressibility evolves smoothly (without any divergence) within the range of $\nu<0.42$, confirming that the system is always metallic. 
The smallness of $\kappa_c^{-1}$ in low density regime indicates a large density of states, in agreement with what expected from a flat band.
While the inverse compressibility increases significantly upon approaching half-filling, indicating that the system approaches an insulator. 

We clarify the distinction between our work and
 several related numerical studies\cite{Palm_2023,ding2023particlehole} on the Hofstadter-Hubbard model have claimed the existence of skyrmion or spin textured metals on the high magnetic field limit. In those systems, the chiral spin textures arise from the exchange interaction and the Hofstadter band. 
 As the correlated electrons are subjected to a uniform magnetic field in these models, these spin textures still align with the conventional understanding of quantum Hall ferromagnets.\cite{PhysRevB.47.16419,PhysRevLett.76.2153}. However, a crucial distinction between question of interest and Hofstadter models lies in that conduction electrons sense a staggered flux with zero net flux. As a result, there is no Landau level in our setting.

What is the origin of the observed spin chirality? If it was indeed through an emergent SOI scenario described in the theory paper\cite{dong2022chiral}, then we should expect the spin chirality emerges from an FM parent state only when orbital magnetization reaches a threshold value. This is because the onset of spin chirality is a result of competition between SOI energy gain (which is proportional to orbital magnetization) and the spin stiffness energy cost which is finite. 

Below we test whether this is indeed the case. For that, we need to study a phase transition where spin chirality emerges on top of FM phase upon tunning orbital magnetization. In our model, orbital magnetization can be directly tunned by tuning the flux $\phi$, e.g. orbital magnetization vanishes at $\phi=0$ by symmetry, and increases as $\phi$ is tuned away from 0. Therefore, we want to study a transition from ``parent" FM to CM1 phase in Fig.\ref{kappac_and_mkHS} driven by flux $\phi$. However, this is not really a good setting to perform such ``experiment" because, as noted in the paragraph above Eq.\eqref{eq:S^2}, the ``parent" FM state in Fig.\ref{kappac_and_mkHS} is only stabilized near a ``magic" point in parameter space where bands are flattened. Directly tuning $\phi$ there would immediately destabilize the FM phase, thus washing out the phase transition of interest. To enable observing such flux-driven phase transition, we need an FM state that is robust upon tuning $\phi$. Fortunately, this can also be achieved in our model in another regime: the regime of infinite $U$
\footnote{In our numerics, the limit of $U\rightarrow \infty$ is implemented by
imposing a nonlocal constraint of no double occupancy at each site, i.e. $n_{j}=0,1$.}
and low doping, where the FM state is robustly stabilized through the Nagaoka scenario, regardless of the value of $t_2/t_1$ and $\phi$. Therefore, below we study flux-driven onset of spin chirality in this Nagaoka regime.

\begin{figure}
		\includegraphics[scale=0.5]{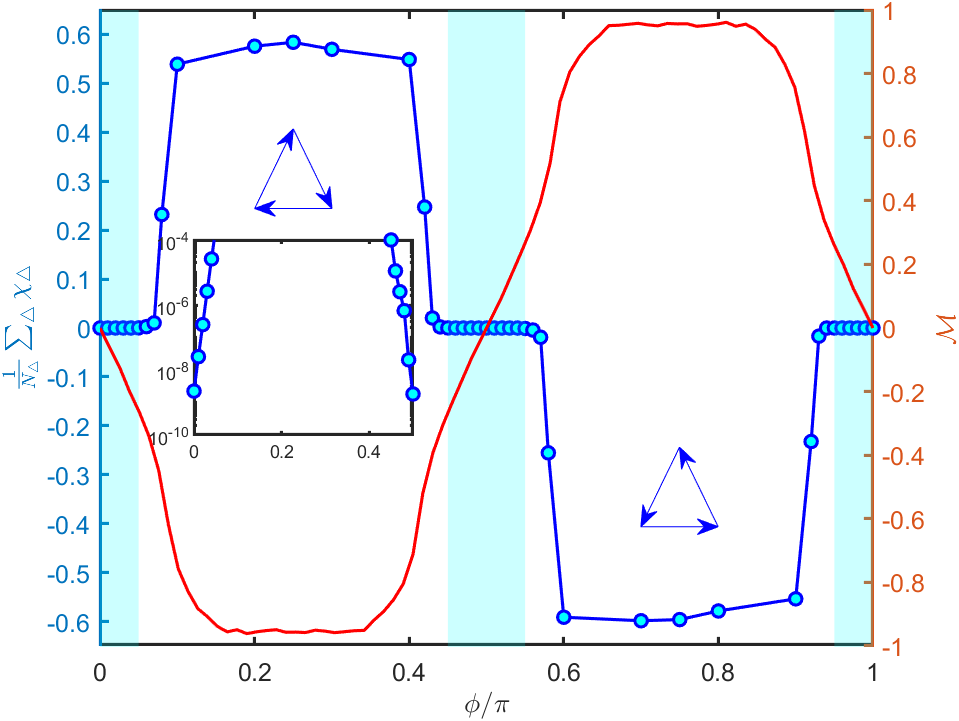}
		\caption{
  (a)At the filling factor of $\nu=29/60$ (two holes away from half-filling), $t_2/t_1=0.2, U=\infty$, the mean spin chirality (blue-dotted line) is plotted as a function of the phase factor $\phi$. The green shadow region represents the zero-spin chirality region. The corresponding total OM (red-line) of the non-interacting system at $\mu=-1$ is also shown(only the lower band is nearly fully occupied). The inset illustrates the semi-log plot $\chi$ versus $\phi$. 
  }
		\label{infUcorr} 
	\end{figure}

The evolution of spin chirality in this flux-driven phase transition is shown in Fig.\ref{infUcorr}. 
Here, we see spin chirality is nonvanishing for generic values of $\phi$, but vanishes inside a finite-width ``plateau" around $\phi=0,\pi/2,\pi$ (colored regime in Fig.\ref{infUcorr}). corresponding to collinear FM orders. These three points are where time-reversal symmetry is restored and orbital magnetization strictly vanishes. This observation matches the expectation that the chirality order should set in at a finite threshold value of $\mathcal{M}$ where the chirality's energy gain exceeds the finite stiffness energy cost.
We note parenthetically that our numerics shows the spin chirality is not strictly zero in this whole plateau (it does not extrapolate to zero under finite-size analysis). 
In fact, from the inset in Fig.\ref{infUcorr}, it decreases exponentially upon decreasing $\phi$ from the edge of the plateau (near $\phi = 0.06\pi$). We attribute this exponentially small chirality to quantum fluctuation that is not captured by classical description in Ref.\cite{dong2022chiral}.

 In addition, we find the plateau of approximate zero spin chirality is broadened on cylindrical geometry where the boundary condition at $y=0$ is set to be periodic (see Supplement \footnote{See more results on cylinder geometry, marked with ``$y$-PBC''.} ). This is also in line with expectation because, enclosing the boundary along $y$-direction introduces a ``frustration" for spin texture\footnote{A simple way to see it is by considering the extreme limit of a two-leg ladder, where introducing PBC directly enforces the chirality in two nearby plaquettes along the $y$ direction to cancel each other out.}, which raises the energy cost of spin chirality thus widening the plateau.

The spin chirality that we focused on in our analysis is not directly observable in experiments.
However, we emphasize that what we found numerically is still meaningful as there are observable signatures that are linked to the spin chirality, such as spin polarization. 
 As the charge carriers with the majority or minority spin 
 sense opposite magnetic fields, the energies of majority-spin and minority-spin carriers are shifted downward and upward by 
the emergent SOI, respectively. 
Namely, the amount of the energies shift between spin-majority and spin-minority carriers at the Fermi level is given by $\delta E_\sigma(k) = \sigma B m_\sigma(k)$. 
This would result in a spatial modulation of spin polarization that mirrors the pattern of spin chirality, if spin chirality is nonuniform, 
\begin{equation}
    \begin{aligned}
        \delta n_{\sigma}(\boldsymbol{r})=&\sigma B(\boldsymbol{r}) \oint_{k_F^\sigma}\frac{{m_{\sigma}(\vec k)d\vec k}}{(2\pi)^2 v_F^\sigma(\vec k)}\\
    \end{aligned}
    \label{polarization_estimation}
\end{equation}
where the integration of $\boldsymbol{k}$ is along the Fermi surface. Indices $\sigma=+,-$ represent the carriers of majority and minority spin, respectively. ${v}_F^{\sigma}(\vec k )$ and $m_{\sigma}(\vec k)$ refer to the magnitude of Fermi velocity and the self-rotation orbital magnetic moment at momentum $\vec k$, respectively. 

In summary we studied a staggered-flux Hubbard model, which is one minimal model that hosts itinerant ferromagnetism with band Berry curvature. Starting from a normal ferromagnet and increasing doping, we find a cascade of phase transitions due to the competition between spin chirality and ferromagnetic order. It first enter a phase with both spin chirality and ferromagnetic order, and then enters another phase where spin chirality suppresses ferromagnetic order.
By scanning the flux in our model, we find that the onset of spin chirality occurs at a finite threshold value of orbital magnetization, which is in line with the theoretical prediction \cite{dong2022chiral}.

	\begin{acknowledgments}
We thank Leonid Levitov, Ting-Kuo Lee and Zhihuan Dong for their fruitful suggestions.  S.Y. and Y. C. are supported by the National Key Research and Development Program of China Grant No. 2022YFA1404204, and the National Natural Science Foundation of China Grant No. 12274086. Z. D. acknowledge the support of the Gordon and Betty Moore Foundation’s EPiQS Initiative, Grant GBMF8682.
		
	\end{acknowledgments}

	
	
	\bibliography{Chiral_Spin_Textures_SOI.bib}


\clearpage

\startsupplement

\begin{widetext}

\begin{center}
\textbf{Supplemental Material for ``Numerical evidence for spin chirality emerging from itinerant ferromagnets in bands
with Berry curvature"}    
\end{center}

\title{Supplemental Material for ``Numerical evidence for spin chirality emerging from itinerant ferromagnets in bands
with Berry curvature"}

\author{Shuai Yang}
\affiliation{Department of Physics and State Key Laboratory of Surface Physics, Fudan University, Shanghai 200433, P.R. China}
\author{Zhiyu Dong}
\affiliation{Department of Physics and Institute for Quantum Information and Matter, California Institute of Technology, Pasadena, California 91125, USA}
\affiliation{Department of Physics, Massachusetts Institute of Technology, Cambridge, MA 02139, USA}
\author{Yan Chen}
\email{yanchen99@fudan.edu.cn}
\affiliation{Department of Physics and State Key Laboratory of Surface Physics, Fudan University, Shanghai 200433, P.R. China}
\affiliation{Shanghai Branch, Hefei National Laboratory, Shanghai 201315, P.R. China}
\date{\today}


This supplementary material contains the process and results of calculating the orbital magnetization of single-particle Hamiltonian, as well as more results of DMRG numerical calculations, including parameter settings for convergence, measurements of the ground state under several characteristic parameters, and results on the cylindrical geometry(compare with the results on open cluster).
\section{Band structures and orbital magnetization for the single particle model}
To investigate the orbital magnetization properties of the model, we invoke a well-known formula based on the semi-classical wave packet paradigm for Bloch electrons\cite{PhysRevB.53.7010} as shown in the following.
\begin{equation}
		\begin{aligned}	\mathcal{M}(\mu)=&\sum_{n}\int^{\mu}\frac{d\boldsymbol{k}}{(2\pi)^2}[m_n(\boldsymbol{k})+\frac{e}{\hbar}\Omega_n(\boldsymbol{k})[\mu-\varepsilon_n(\boldsymbol{k})]]\\=&\mathcal{M}_c+\mathcal{M}_\Omega
		\end{aligned}
		\label{om_formula}
	\end{equation} 
Here $\varepsilon_n(\boldsymbol{k})$ is the band structure for the non-interacting part of this model. The term $m_n(\boldsymbol{k})$ in Eq(\ref{om_formula}) represents the orbital magnetic moment generated by self-rotation of electron wave packets at momentum $\boldsymbol{k}$, taking the form of $m_n(\boldsymbol{k})=-\frac{ie}{2\hbar}\left<\nabla_{\boldsymbol{k}}u_n(\boldsymbol{k})\mid\times[H(\boldsymbol{k})-\varepsilon_n(\boldsymbol{k})]\mid \nabla_{\boldsymbol{k}}u_n(\boldsymbol{k})\right>$, in which $H(\boldsymbol{k})$ is the crystal Hamiltonian. 
	The Berry curvature	$\Omega_n(\boldsymbol{k})=i\left<\nabla_{\boldsymbol{k}}u_n(\boldsymbol{k}) \mid\times \mid\nabla_{\boldsymbol{k}}u_n(\boldsymbol{k})\right>$
	, also contributes to the orbital magnetization due to the center-of-mass motion of electron wave packets. The integration runs over states with energies below the Fermi energy $\mu$, the first part in Eq(\ref{om_formula}) is the conventional part $\mathcal{M}_c$ arising from self-rotation of electron wave pocket and the second part $\mathcal{M}_{\Omega}$ accounts for the contribution from the chiral edge state of a topological band. 
 
But directly calculating $\Omega_n(\boldsymbol{k})$ and ${m}_n(\boldsymbol{k})$ for this model is complicated. Therefore, we first insert a complete base $1=\sum_{n'}\left|u_{n'}\right>\left<u_{n'}\right|$ and rewrite the formula as follows:
\begin{equation}
    \begin{aligned}
        &\Omega_n(\boldsymbol{k})=i\sum_{n'\ne n}{\left<\frac{\partial u_n}{\partial k_x}\mid u_{n'}\right>\left<u_{n'}\mid\frac{\partial u_n}{\partial k_y}\right>-c.c}\\
        &m_n(\boldsymbol{k})=\frac{ie}{\hbar}\sum_{n'\ne n}{\left<\frac{\partial u_n}{\partial k_x}\mid H-\varepsilon_n\mid u_{n'}\right>\left<u_{n'}\mid\frac{\partial u_n}{\partial k_y}\right>-c.c}
    \end{aligned}
\end{equation}
Further with the Identity as:
\begin{equation}
    \left<u_{n'}\mid\frac{\partial}{\partial \boldsymbol{k}}\mid u_n\right>=\frac{\left<u_{n'}\mid\frac{\partial H(\boldsymbol{k})}{\partial \boldsymbol{k}}\mid u_n\right>}{\varepsilon_{n'}-\varepsilon_n}
\end{equation}

The Berry curvature and the orbital magnetic moment carried by the Bloch electron can be rewritten in the following form:
\begin{equation}
    \begin{aligned}
        \Omega_n(\boldsymbol{k})&=i\sum_{n'\ne n}\frac{\left<u_n\mid \frac{\partial H}{\partial k_x}\mid u_{n'}\right>\left<u_{n'}\mid\frac{\partial H}{\partial k_y}\mid u_n\right>}{\left(\varepsilon_{n'}-\varepsilon_n\right)^2}-c.c.
\\
        m_n(\boldsymbol{k})&=i\frac{e}{\hbar}\sum_{n'\ne n}\frac{\left<u_n\mid \frac{\partial H}{\partial k_x}\mid u_{n'}\right>\left<u_{n'}\mid\frac{\partial H}{\partial k_y}\mid u_n\right>}{\varepsilon_{n'}-\varepsilon_n}-c.c.
    \end{aligned}
\end{equation}
which is more tractable for numerical calculation.

The Fig\ref{Ek_and_flat} below shows the band dispersion relationship $\varepsilon_n(\boldsymbol{k})$((a),(d)), Berry curvature $\Omega_n(\boldsymbol{k})$((b),(e)), and orbital magnetic moment $m_n(\boldsymbol{k})$((c),(f)) information distributed along highly symmetric points under three different characteristic parameters.
\begin{figure}
		\includegraphics[scale=0.2]{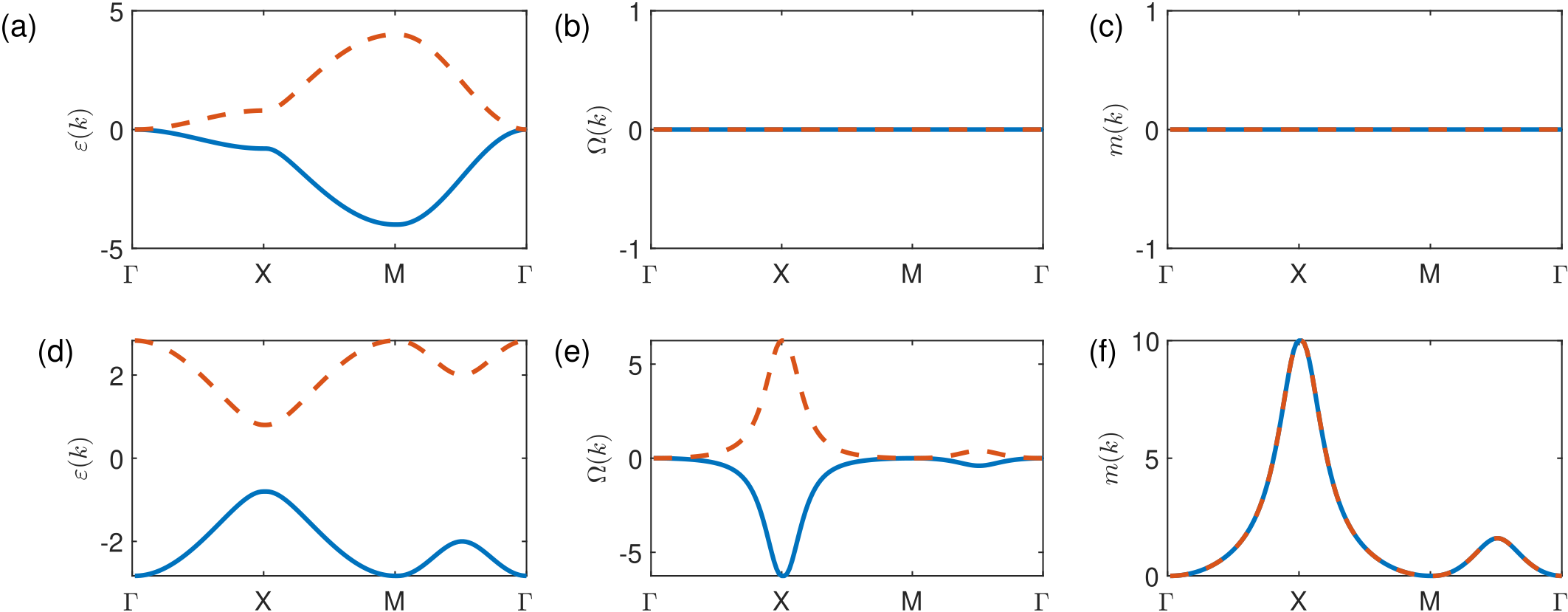}
		\caption{ At $\phi=\pi/4$, the band structure of the tight binding model at (a) $t_2/t_1=0.0$, and (d) $t_2/t_1=0.2$. The Berry curvature $\Omega_n(\boldsymbol{k})$ of (b) $t_2/t_1=0.0$, and (e) $t_2/t_1=0.2$. The orbital magnetic moment $m_n(\boldsymbol{k})$ of (c) $t_2/t_1=0.0$, and (f) $t_2/t_1=0.2$
  }
		\label{Ek_and_flat} 
	\end{figure}
At $\phi=0,\pi/2,\pi$, Berry curvature and orbital magnetic moment are both equal to zero, due to the symmetry of the time reversal and the space inversion.
At $\phi=\pi/4$, the non-zero $t_2$ causes the system to open the gap at $X(\pi,0)$, and at the same time breaks the symmetry of time reversal and space inversion, thus obtaining the non-vanishing Berry curvature and orbital magnetic moment.

The Fig\ref{M_mu_phi} below illustrates the two components of the total orbital magnetization calculated in the main text. It can be seen that the OM from these two parts are always opposite, reflecting the opposite circulation motion of the two components. All these results are consistent with the previous articles\cite{PhysRevB.76.064406,PhysRevB.76.094406,PhysRevLett.95.137205,PhysRevB.74.024408}.
\begin{figure}
		\includegraphics[scale=0.65]{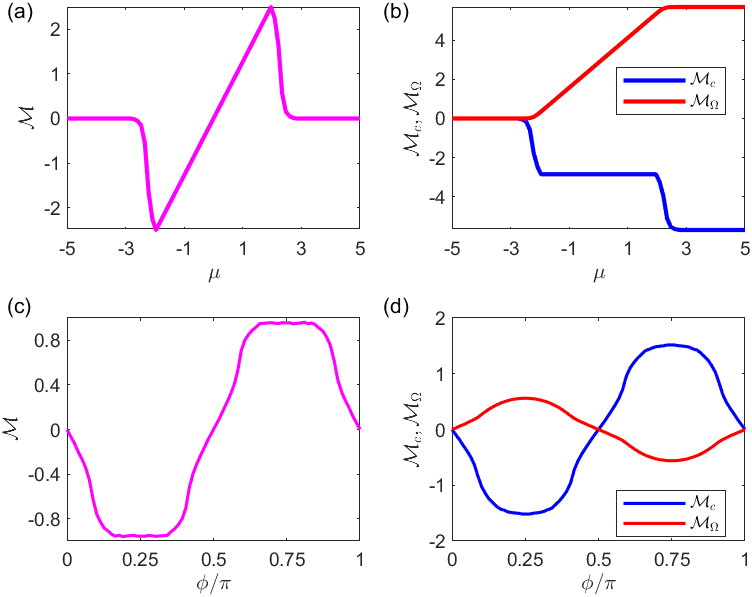}
		\caption{Orbital magnetization of the non-interacting model as a function of (a)chemical potential $\mu$ and (c)phase factor $\phi$. (b)Its two components $\mathcal{M}_c$ (blue line) and $\mathcal{M}_\Omega$ (red line) as a function of electron chemical potential $\mu$ at $t_2/t_1=0.6,\phi=\pi/4$. (d)$\mathcal{M}_c$ (blue line) and $\mathcal{M}_\Omega$ (red line) as a function of phase factor $\phi$ at $t_2/t_1=0.2,\mu=-1$. 
  }
		\label{M_mu_phi} 
	\end{figure}  
\section{Additional numerical simulation information}
\subsection{Details of the numerical simulation}
We perform DMRG calculations\cite{itensor} for this generalized Hubbard model on multi-leg ladders (both open boundary conditions in $x$ and $y$ directions) and multi-leg cylinders(open boundary condition in $x$ direction and periodic boundary condition in $y$ direction). 
We mainly compute the properties of 4-leg and 6-leg systems (with sizes up to $4\times 20$ and $6\times 10$. For the $4\times10$ system, the maximum truncation error is approximately in the order of $10^{-5}$.
For the $4\times20$ and $6\times10$ systems, the maximum truncation error is approximately $10^{-4}$. However, there are also some parameter regimes where the attained truncation error is around $10^{-6}$.). We perform at most $200$ sweeps to search for the ground state within the sector of the total $S_z$ component equal to zero and the bond dimension is set up to $D=8000$ to get converged results. We believe these results are reliable in predicting the behavior of electrons in realistic two-dimensional systems. 
\subsection{Convergence of the DMRG calculation}
In this part, we show the bond dimension dependence of ground state energy per site $E_0/N_s$(Fig.\ref{eg_svn_chi_D}(a)), entanglement entropy $S_E$(Fig.\ref{eg_svn_chi_D}(b)) and mean spin chirality $\chi_\triangle$(\ref{eg_svn_chi_D}(c)) for the parameter corresponding to the inset of Fig.\ref{kappac_and_mkHS}(b) in the main text. During the calculation, the bond dimension is set as $D=3000,4000,5000,6000,7000,8000$. For these metallic states, we found that the entanglement entropy is always rapidly increasing, indicating that these states are challenging for DMRG calculations. However, the results for energy and average spin chirality show relatively good convergence.

\begin{figure}
		\includegraphics[scale=0.6]{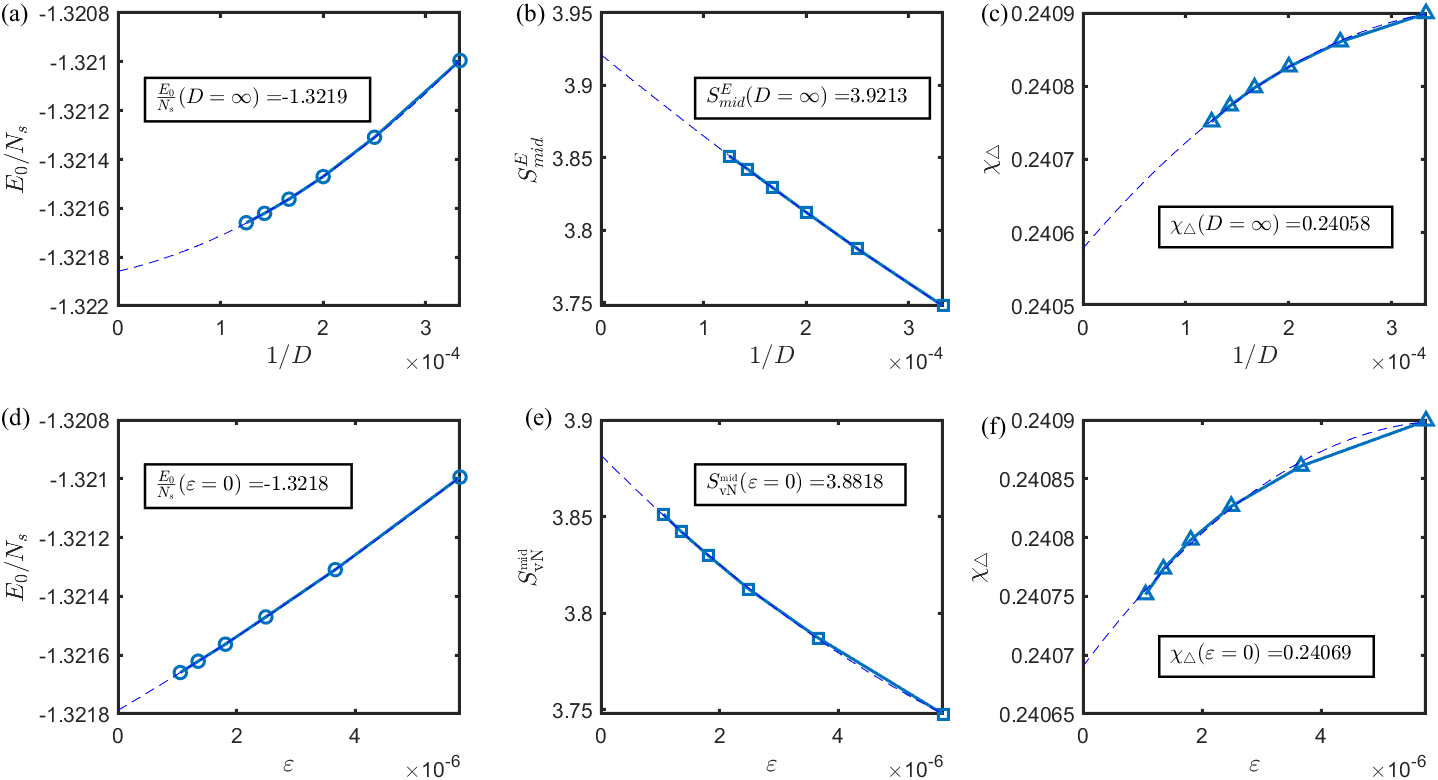}
		\caption{(a)The ground state energy per site $E_0/N_s$, (b)Entanglement entropy $S_{mid}^E$ measured at the center of the systems, (c) mean spin chirality $\chi_\triangle$ as a function of the inverse of bond dimension $1/D$. (d)(e)(f) same data versus truncation error $\varepsilon$. Using Second-order polynomial extrapolation (as indicated by the dashed line) to obtain the result at $D=\infty$ or equivalently $\varepsilon=0$ (marked with a text box in the corresponding graph).}
		\label{eg_svn_chi_D} 
	\end{figure} 

Additionally, the results below(see Fig.\ref{corr_D}) show the single-particle correlation $\left|\left<c_{0\sigma}^\dagger c_{r\sigma}\right>\right|$, spin-spin correlation$\left|\left<\boldsymbol{S_0}\cdot\boldsymbol{S_r}\right>\right|$, and chiral-chiral correlation$\left|\left<\chi_{ 0}\chi_{ r}\right>\right|$ for different bond dimensions, further confirming that the ground state is an incompressible metal with spin chirality. The square root of the chiral-chiral correlation at the largest distance $\sqrt{\chi_{ 0}\chi_{R}}\approx 0.23612$ ($R=10$) is also close to the average spin chirality $\chi_\triangle(D=\infty)\approx 0.24058$ measured for a single triangle.

\begin{figure}
		\includegraphics[scale=0.5]{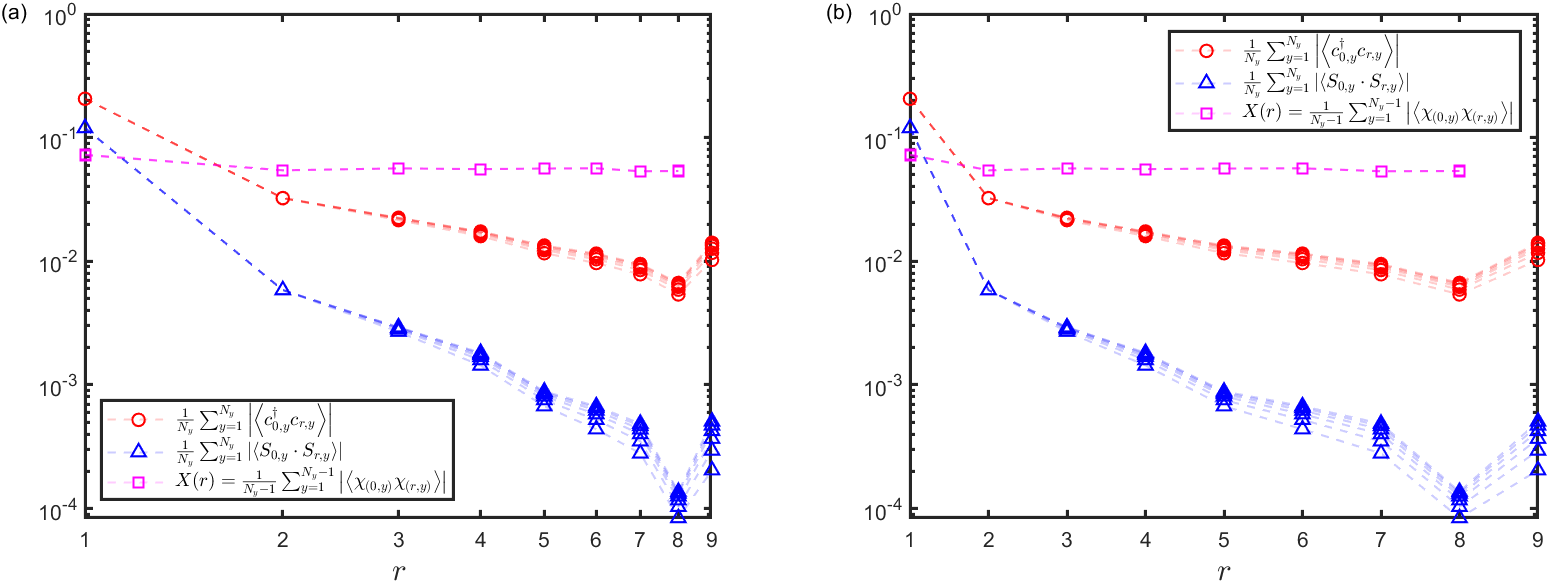}
		\caption{(a)Log-log plot (b) and Log-linear plot of single-particle Green's function $\left|\left<c_{0\sigma}^\dagger c_{r\sigma}\right>\right|$, spin-spin correlation $\left|\left<\boldsymbol{S}_0\cdot\boldsymbol{S}_r\right>\right|$, and chiral-chiral correlation $\left|\left<\chi_0\chi_r\right>\right|$ versus distance $r$, which labels the column index. The dashed line corresponds to the results for different bond dimensions($D=3000,4000,5000,6000,7000,8000$).}
		\label{corr_D} 
	\end{figure} 

 \subsection{the magnetization in chiral metal phase}
 The Fig. \ref{fig:power_and_length} below shows the evolution of the magnetization $S^2/S_{\text{max}}^2$ as a function of the filling factor. In the FM and CM1 phases, $S^2/S_{\text{max}}^2$ remains a finite value, indicating that the system possesses magnetic long-range order in the thermodynamic limit. In contrast, the CM2 phase does not exhibit magnetic long-range order.

 \begin{figure}
     \centering
     \includegraphics[width=0.5\linewidth]{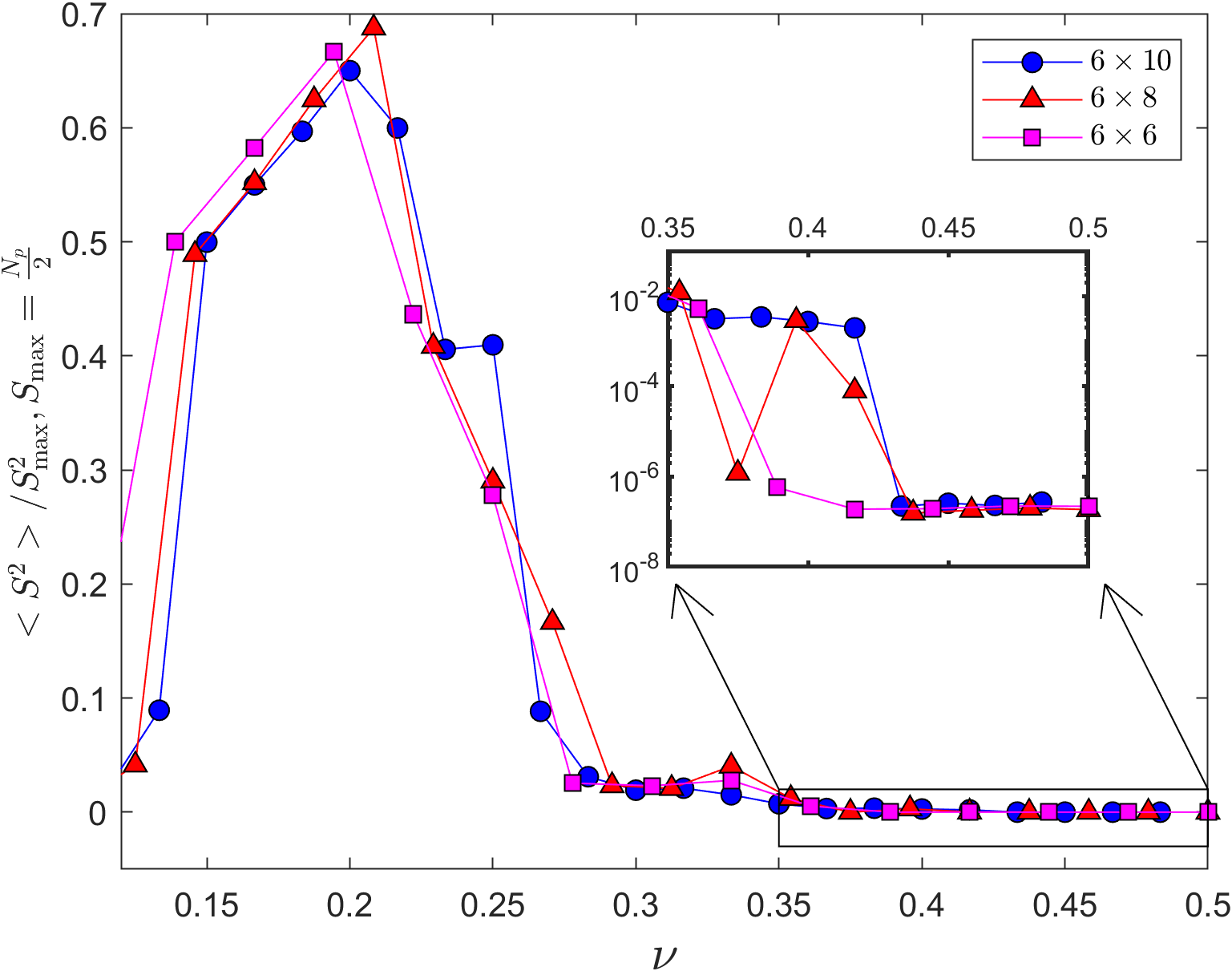}
     \caption{The filling dependence of total magnetization.}
     \label{fig:power_and_length}
 \end{figure}
\subsection{Results for the y-PBC cylinder geometry}
In this part, we present numerical simulation results on cylindrical geometry to further support our argument. The results on the cylinder and the results with open boundary conditions in both $x$ and $y$ directions exhibit similar trends. So the conclusions are roughly consistent. The main difference is the presence of larger zero-spin chiral regions in cylindrical geometry, as we explained in the main text.

From Fig.\ref{chiral_as_phi_and_mu_OPBC}(a), when adjusting the particle population density, the system with periodic boundary condition in the $y$ direction has a certain lag in entering the chiral metal phase. The same phenomenon is observed when adjusting the phase factor attached to the electron hopping. As illustrated in Fig\ref{chiral_as_phi_and_mu_OPBC}(b), the spin chirality is nearly zero when approaching the symmetric points of $\phi=0,\pi/2,\pi$. Moreover, the results on the cylinder show that there is a larger parameter interval around these symmetric points to keep the system in the ferromagnetic phase.

\begin{figure}
		\includegraphics[scale=0.5]{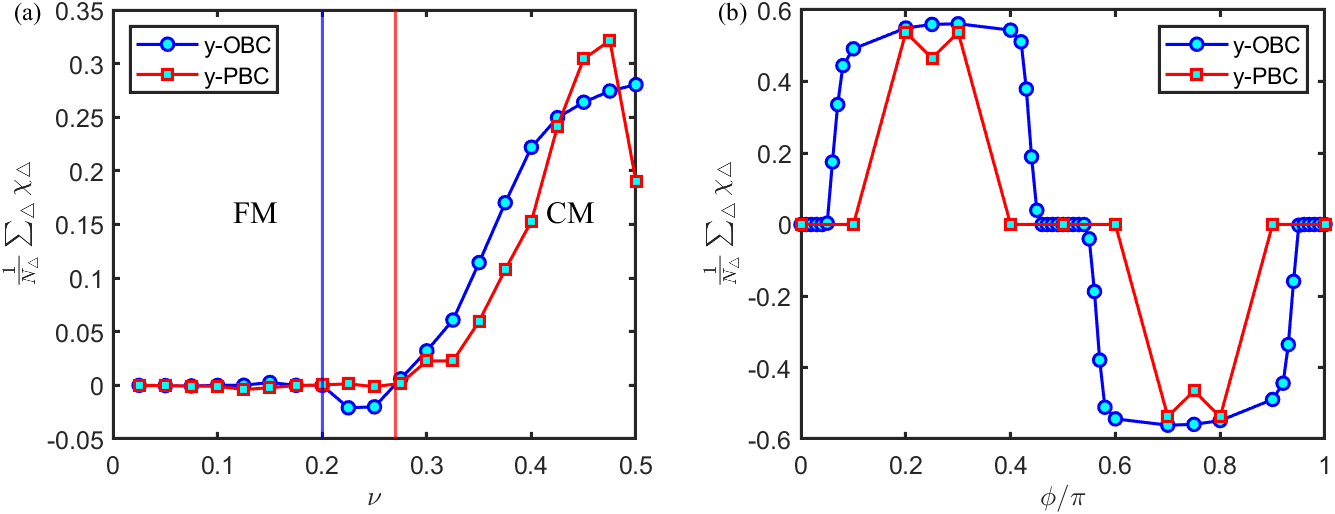}
		\caption{Mean chiral spin order as a function of different (a) filling factor $\nu$ with $U=4,t_2/t_1=0.6$, (b)and $\phi/\pi$ with $U=\infty,t_2/t_1=0.2$ on $N_x\times N_y=10\times 4$ open cluster(depicted with blue-circle) and cylinder (depicted with red-square).
		}\label{chiral_as_phi_and_mu_OPBC} 
	\end{figure}

\end{widetext}
\end{document}